# DNA-TO-GO: A PORTABLE SMARTPHONE-ENABLED PCR ASSAY PLATFORM

A. Priye and V.M. Ugaz

*Department of Chemical Engineering, Texas A&M University, College Station, TX 77843, USA*

**ABSTRACT**

We address the need for affordable, rapid, and easy to use diagnostic technologies by coupling an innovative thermocycling system that harnesses natural convection to perform rapid DNA amplification via the polymerase chain reaction (PCR) with smartphone-based detection. Our approach offers an inherently simple design that enables PCR to be completed in 10-20 minutes. Electrical power requirements are dramatically reduced by harnessing natural convection to actuate the reaction, allowing the entire system to be operated from a standard USB connection (5 V) via solar battery packs. Instantaneous detection and analysis are enabled using an ordinary smartphone camera and dedicated app interface.

**KEYWORDS:** Convection, PCR, Smartphone, DNA analysis, Point of care, Detection

**INTRODUCTION**

The polymerase chain reaction (PCR) has become a routine molecular biology technique for a variety of applications such as reliable detection and diagnosis of infectious diseases. PCR requires periodic heating and cooling of the reagents, which is conventionally performed by thermocycling instruments that repeatedly change the temperature of high thermal mass blocks—a format that consumes considerable electrical power and is not adapted for portability. Successful DNA amplification is then quantified either in real time via fluorescence detection or after PCR through gel electrophoresis requiring additional instrumentation and time. With recent advances in microfluidics, there have been constant efforts geared towards developing PCR systems that are simpler, less expensive, fast, and portable. Expanded availability of such technologies will not only impact health care, but will be a key to enabling rapid identification of infectious agents in environments with little or no laboratory infrastructure.

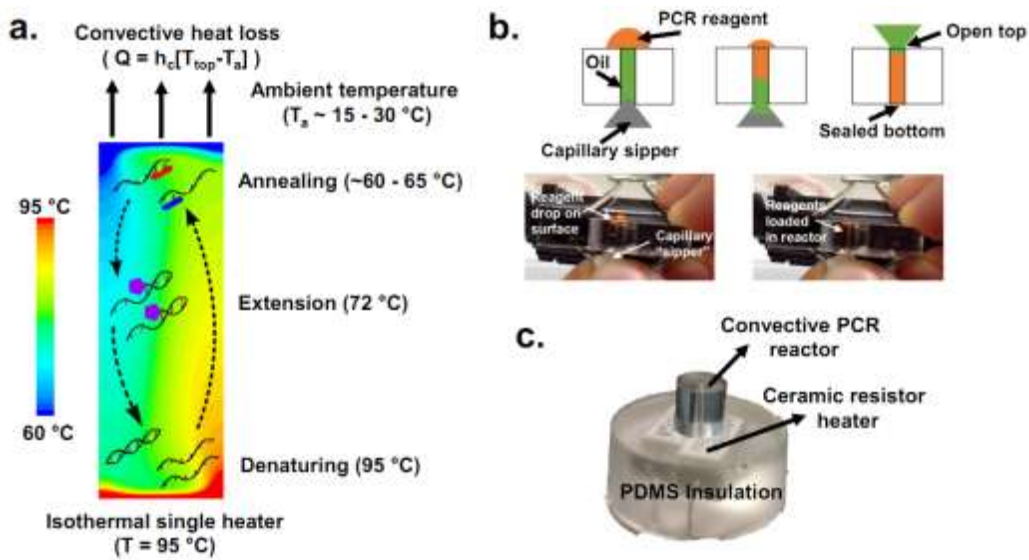

*Figure 1: (a) Rayleigh Bernard convection enables PCR to be performed in a pseudo-isothermal manner with denaturing, annealing and extension reactions occurring at different temperature zones within the reactor. The top surface experiences heat loss (Q) to the surroundings ($T_a$) at a rate dependent on the convective heat transfer coefficient ($h_c$). (b) A capillary sipper mechanism ensures bubble free loading of the PCR reagents in the reactor ensuring an airtight seal. (c) The PCR reactor is mounted on a single ceramic heater encased in polydimethylsiloxane (PDMS) for insulation.*

## THEORY

We have devised a simple Rayleigh-Bernard convection based thermocycler for rapid DNA amplification coupled with smartphone imaging for real time fluorescence detection of reaction products. The PCR solution is enclosed in a cylindrical volume heated from below. The sealed top surface is cooled by convective heat loss to the environment (Fig. 1a). The imposed vertical temperature gradient induces natural convection, generating intricate convective flow fields which can uniquely be mapped by parameters like the cylinder dimensions (height, $h$ and diameter, $d$) and thermal conditions (Rayleigh number, $Ra$). The cylinder geometry can then be tuned to generate complex 3-D chaotic flow fields, conducive to robust convective PCR [1-8]. The simple idea makes it possible to mount micro cylindrical PCR reactors on a single heater which maintains the bottom temperature at 95 °C and sustains a convective flow capable of automatically transporting the reagents from the hot bottom region where denaturing takes place to the cooler top region where annealing and extension takes place thus enabling rapid convective PCR in a pseudo-isothermal manner.

## EXPERIMENTAL

Reagent loading is achieved using a novel capillary-based approach that is self-sealing and requires minimal manual sample handling. The operator need only pace a drop of the PCR reagents on the outer surface of a disposable reactor cartridge, after which the fluid is passively drawn into the reactor with the aid of a capillary "sipper" device that also seals the reagents inside the chamber by leaving a thin layer of oil on top (Fig. 1b). Low power operation is enabled since the bottom heater functions in an isothermal manner (Fig. 1c). Thermal management is achieved using an off-the-shelf ceramic resistor with a polydimethylsiloxane (PDMS) casing for insulation, interfaced with an Arduino microcontroller board programmed with a MOSFET-based temperature feedback loop (Fig. 2a). This design dramatically reduces electrical consumption, enabling the entire reaction to be performed within the power limitations of a conventional 5 V USB source such as solar battery packs.

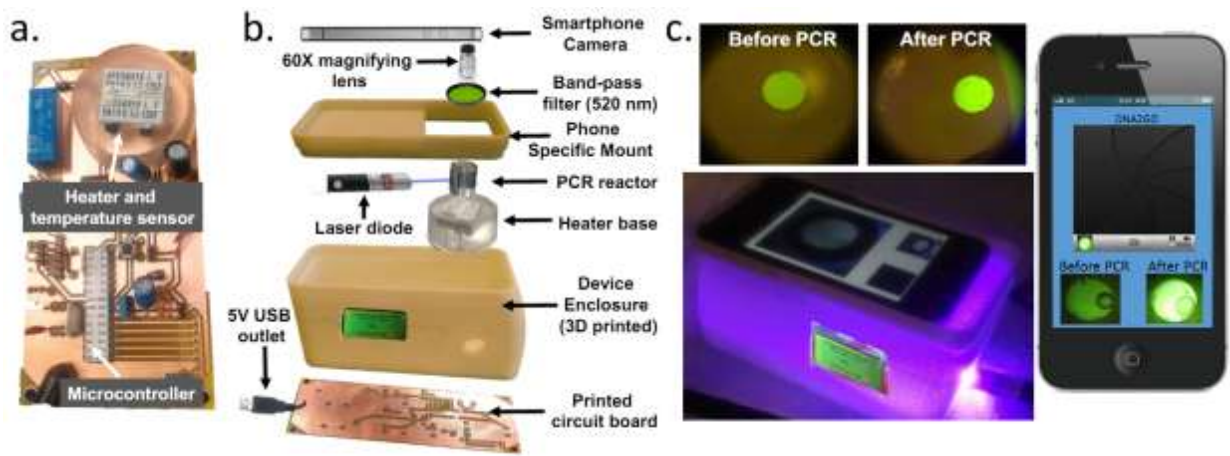

*Figure 2: (a) Circuit board consisting of a ceramic heater, temperature sensor, relay for laser diode and a microcontroller programmed via Arduino interface. (b) Assembly of USB-powered DNA analysis device showcasing different components including the heater and laser pointer diode for sample illumination. (c) The assembled device in operation. PCR products are detected and analyzed using a smartphone camera and app.*

Fluorescence detection is achieved by incorporating a diode obtained from a blue laser pointer (460 nm, 100 mW) coupled with an excitation (480 nm) and emission (520 nm) band-pass filter set that permits an ordinary smart phone camera to detect successful PCR replication using a SYBR-Green based dye chemistry (Fig. 2b). A dedicated smartphone app enables acquisition and quantitative analysis of the fluorescence images, including background subtraction from negative control reactions and intensity normalization ($I = [I_{measured} - I_{min}]/[I_{max} - I_{min}]$) (Fig. 2c). Detection can be performed either in end-point mode or to monitor

DNA replication in real time during the course of the reaction. Results are typically obtained in approximately 10 - 20 minutes of operation.

All components are integrated and housed in an enclosure incorporating an interchangeable smartphone cradle that ensures proper alignment of the camera and optical components regardless of the type of mobile device employed (Fig. 2c).

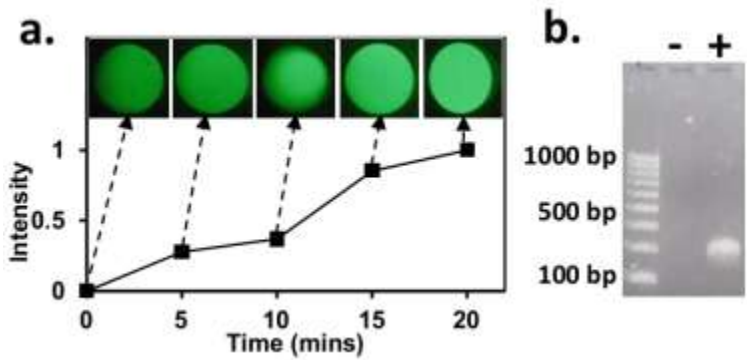

*Figure 3: Amplification of a 237 bp sequence of λ DNA (initial concentration = 1 μg/mL) with SYBR green chemistry for fluorescence detection in a cylindrical reactor (h = 10 mm, d = 2mm). (a) The iPhone app enabled real time image acquisition of the top surface at regular time intervals. An algorithm was then used to subtract background fluorescence and normalize the intensity of each image.(b) Gel electropherogram displaying negative control run (without DNA) and positive run (with DNA) confirming amplification of correct sequence in convective PCR.*

## RESULTS
The device was tested by replicating a 237 base pair target from λ-phage DNA template in a cylindrical reactor with $h$ = 10 mm and $d$ = 2 mm. A typical 100 μL reaction mixture contained 50 μL of SYBR master mix, 10 μL of each reverse and forward primers (10 μM), 1 μL of λ-phage DNA (100 μg/mL) and 29 μL of PCR grade water. The iPhone app was run in continuous mode to capture images of reactor top every five minutes. An increase in fluorescence intensity corresponds to a positive PCR run and the amplification was also verified by detection on the gel (Fig. 3).

## CONCLUSION
By combining the benefits and speed of the convective PCR format with the versatility of a smart phone-based detection platform it is possible to construct a complete DNA analysis system for approximately $20 ($US). The ability to deliver performance comparable to or surpassing that of current-generation systems while simultaneously providing an orders of magnitude reduction in cost has potential to greatly expand the use of PCR-based detection assays by moving them out of the laboratory and into settings where they are needed most. The smart phone-based platform can also be leveraged for distributed operation by transmitting test results including sample identification information, field notes, and GPS coordinate tags to a centralized database for archival and analysis.

## ACKNOWLEDGEMENTS
We gratefully acknowledge involvement by a team of 14 undergraduate students who participated in this project as part of the AggiE-Challenge program in the College of Engineering at Texas A&M University.

## REFERENCES

[1] Priye, Aashish, et al "Lab-on-a-drone: toward pinpoint deployment of smartphone-enabled nucleic acid-based diagnostics for mobile health care". Analytical chemistry 88. (2016) doi: 10.1021/acs. analchem.5b04153
[2] Priye, Aashish, Yassin A. Hassan, and Victor M. Ugaz. "Microscale chaotic advection enables robust convective DNA replication." Analytical chemistry 85.21 (2013): 10536-10541.
[3] Priye, Aashish, Yassin A. Hassan, and Victor M. Ugaz. "Education: DNA replication using microscale natural convection." Lab on a Chip 12.23 (2012): 4946-4954.

[4] Priye, Aashish, Yassin A. Hassan, and Victor M. Ugaz. "Selecting 3D Chaotic Flow States for Accelerated DNA Replication in Micro Scale Convective PCR" The 15th International Conference on Miniaturized Systems for Chemistry and Life Sciences (2011)

[5] Priye, Aashish, and Victor M. Ugaz. "Convective PCR Thermocycling with Smartphone-Based Detection: A Versatile Platform for Rapid, Inexpensive, and Robust Mobile Diagnostics." Microfluidic Methods for Molecular Biology. Springer International Publishing, (2016). 55-69.

[6] Priye, Aashish, S. Wong , and Victor M. Ugaz. "Lab-On-A-Drone Deployment Of Nucleic Acid-Based Diagnostics" The 19th International Conference on Miniaturized Systems for Chemistry and Life Sciences (2015)

[7] Priye, Aashish, and Victor M. Ugaz. "Microscale Chaotic Advection Enables Enhanced Surface Electrochemistry in Hydrothermal Pore Environments" The 19th International Conference on Miniaturized Systems for Chemistry and Life Sciences (2015)

[8] Priye, Aashish, and Victor M. Ugaz. " Thermally-Targeted Adsorption And Enrichment In Micro-scale Hydrothermal Pore Environments" The 17th International Conference on Miniaturized Systems for Chemistry and Life Sciences (2013)